\documentclass[sigplan,nonacm]{acmart}
\setcopyright{none}
\acmConference{}{}{}
\acmDOI{}
\acmISBN{}
\date{}
\AtBeginDocument{}

\hypersetup{pdfauthor={Bangbo Liang, Yupeng Chen, Sicheng Zhao, Peihao Huang, Di Yang, Bohua Xu, Bin Yang, Shizhen Zhao, Guo Chen}}

\begin{document}
\title[LACE: Lane-Granularity OCS Scheduling]{Breaking the Duplex Barrier: Lane-Granularity OCS Scheduling for LLM Training}
\settopmatter{authorsperrow=3}
\author{Bangbo Liang}
\email{lbb@hnu.edu.cn}
\affiliation{
  \institution{Hunan University}
  \country{China}
}
\author{Yupeng Chen}
\email{cyp0633@hnu.edu.cn}
\affiliation{
  \institution{Hunan University}
  \country{China}
}
\author{Sicheng Zhao}
\email{zhaosicheng@hnu.edu.cn}
\affiliation{
  \institution{Hunan University}
  \country{China}
}
\author{Peihao Huang}
\email{hph101001@hnu.edu.cn}
\authornote{Corresponding authors.}
\affiliation{
  \institution{Hunan University}
  \country{China}
}
\author{Di Yang}
\email{yangd18@chinaunicom.cn}
\affiliation{
  \institution{China Unicom Software Research Institute}
  \country{China}
}
\author{Bohua Xu}
\email{xubh15@chinaunicom.cn}
\affiliation{
  \institution{China Unicom Research Institute}
  \country{China}
}
\author{Bin Yang}
\email{researcher\_yang@outlook.com}
\affiliation{
  \institution{China Unicom Research Institute}
  \country{China}
}
\author{Shizhen Zhao}
\email{shizhenzhao@sjtu.edu.cn}
\affiliation{
  \institution{Shanghai Jiao Tong University}
  \country{China}
}
\author{Guo Chen}
\email{guochen@hnu.edu.cn}
\authornotemark[1]
\affiliation{
  \institution{Hunan University}
  \country{China}
}
\renewcommand{\shortauthors}{Liang et al.}

\begin{abstract}
An optical circuit switch (OCS) can reconfigure physical connectivity to match the predictable communication schedules of large language model (LLM) training.
Although each OCS light path is physically simplex, existing demand-aware OCS schedulers allocate capacity in duplex-port pairs, forcing equal bandwidth in both directions and stranding capacity under asymmetric node-pair traffic.

This paper presents \textbf{LACE}, the first offline OCS schedule compiler that independently allocates transmit (TX) and receive (RX) lanes for LLM training.
Without changing the selected collective algorithms, operation order, or rank placement, LACE reconstructs directed node-level demand, jointly determines which consecutive operations share a configuration and how many simplex circuits serve each direction, and realizes these allocations as physical lane bindings and optical paths under per-node lane-inventory and multi-OCS fabric constraints.
Software acknowledgments carry feedback over independently provisioned return paths, while coordinated link configuration and recovery verify each configuration before communication resumes.
On a separate three-server testbed using fixed topologies and matched per-port rate limits, LACE's asymmetric connectivity achieves $1.80\times$ speedup for communication replay and $1.27\times$ for GPT-2 training over a symmetric-topology baseline. At larger scale, simulations of LLaMA-3.1 70B and 405B schedules with sixteen 400-Gb/s ports per server show that LACE achieves $1.21$--$2.04\times$ communication speedup over the latest duplex OCS scheduler.
\end{abstract}

\maketitle

\section{Introduction}
Large-scale LLM training generates enormous inter-node traffic, making network bandwidth a first-order determinant of training performance and infrastructure cost.
 For training jobs with a known communication
schedule, operators can reconfigure the physical fabric to match
communication demand~\cite{topoopt}, rather than relying solely on a
static topology provisioned for worst-case demand. An optical circuit
switch (OCS) fabric, consisting of one or more OCS devices, provides
this flexibility: it steers light between fibers to establish dedicated
paths without intermediate optical-to-electrical conversion. Commercial
MEMS OCSs can change these optical connections on millisecond time
scales~\cite{farrington2011hardware,mori2021highport}. Long studied in the data-center literature~\cite{farrington2010helios,cthrough,osa,rotornet,opera,sirius},
OCSs now underpin production infrastructure, including Google's Jupiter
network~\cite{JupiterEvolving,Apollo} and TPU v4 supercomputer~\cite{tpuv4},
and motivate LLM-specific designs that adapt connectivity to training
communication~\cite{lightwave,sipml,topoopt,mixnet}.

For all their diversity, existing demand-aware OCS systems use the duplex port as the basic unit of topology scheduling. Each node is either a server or an electrical packet switch (EPS) and connects to the OCS fabric through one or more ports, each comprising one transmit (Tx) lane and one receive (Rx) lane carried on separate fibers. The topology scheduler computes a matching over ports: connecting port \(i\) to port \(j\) always instantiates both directions of the circuit at once~\cite{tpuv4,JupiterEvolving,Apollo,lightwave,sipml,topoopt}. Nothing in the optics requires this. The mirrors steer each simplex light path independently, and a “duplex circuit” is merely two simplex cross-connects that happen to be coordinated; schedulers could, in principle, issue them separately. They do not, because of what sits at the other end of the fiber: default link-management mechanisms are designed for a port whose Tx and Rx lanes reach the same peer, and prior measurements show that simplex reconfiguration can trigger link training and failure handling at NICs and switches, causing unintended link drops\cite{lane_zerwas}. In our testbed, however, ConnectX-6 ports operate with different Tx and Rx peers once autonegotiation is disabled and the line rate is fixed. Duplex-port binding therefore stems from conventional endpoint link management rather than a physical constraint of the OCS fabric.

That convention was benign when traffic was roughly symmetric. LLM traffic is not. The bytes a node sends and receives diverge sharply between the two directions: pipeline-parallel stages push activations downstream and pull gradients upstream in disjoint time windows\cite{gpipe,pipedream}; a selected DP
ring can send bulk data to one neighbor and receive it from another. A duplex scheduler, however, can only provision capacity in symmetric pairs: matching node \emph{A} to node \emph{B} buys \emph{A} one lane toward \emph{B} and one lane back from \emph{B}, whether or not \emph{B} has anything to send. The cold direction of each circuit idles while the hot direction saturates—and no matter how cleverly the matching is computed~\cite{tpuv4,lightwave,sipml,topoopt,mixnet}, symmetry itself is never up for negotiation.

This paper argues that the symmetry constraint is not only costly but unnecessary—once we are precise about \emph{where} bidirectional connectivity is actually required. In the scheduling model, the ports attached to the OCS fabric are grouped by node: a GPU server contributes its NIC ports, while an EPS contributes its uplink ports. Workloads specify communication between nodes; they do not require the Tx and Rx lanes of each port to connect to the same peer. A node with \emph{k} ports therefore contributes \emph{k} Tx and \emph{k} Rx lanes that can be allocated independently across peers according to directional demand.

However, building a practical lane-level OCS scheduler presents three challenges. First, nodes must remain operational when a port's TX and RX lanes connect to different peers, and transport feedback must reach the sender. Second, node-pair allocations compete for shared TX and RX inventories; in a multi-OCS fabric, they must also fit the available optical paths. Finally, directional demand changes across operations, while every configuration change incurs optical switching and node recovery delay. The scheduler must jointly decide which operations share a configuration and how their lanes are allocated.

In this paper, we present \textbf{LACE}:
\textbf{La}ne-level \textbf{C}ircuit Sch\textbf{e}d\-ul\-ing. LACE treats TX and RX lanes as independently schedulable simplex resources and compiles an ordered training communication schedule offline. Its Demand Profiler constructs directed node-level demand. The Segment Planner chooses configuration boundaries and fractional simplex circuit allocations, and the Circuit Realizer converts these targets into integer counts within the lane inventories. The Lane Binder assigns physical lanes and optical paths, retaining connections where possible and repairing allocations when internal paths are constrained. Software acknowledgments provide return feedback across ports, while coordinated link configuration and recovery verify connections before transmission resumes. Together, these mechanisms match directional demand while accounting for the cost of changing configurations (\S\ref{sec:design}).

We implement LACE's scheduler and prototype node-side communication and recovery mechanisms using commodity OCS hardware. The code will be made available upon formal publication. With sixteen 400Gbps ports per server, simulations of LLaMA-3.1 70B and 405B schedules show $1.21$--$2.04\times$ communication speedup over ACTINA. Our three-server fixed-topology testbed achieves $1.80\times$ communication-replay and $1.27\times$ GPT-2 training speedups over ACTINA. Separate experiments measure software-ACK overhead and verify configuration recovery; simulation evaluates switch-node aggregation and circuit mapping through a multi-OCS Clos.

In summary, we make the following contributions:
\begin{itemize}

\item We quantify directional demand between nodes in LLM training
and show why fixed symmetric or asymmetric circuit allocations mismatch changing DP/PP traffic (\S\ref{background-and-motivation}).

\item We design LACE to jointly plan shared configurations and
fractional lane allocations, then realize integer simplex circuits
and physical paths within node and fabric constraints
(\S\ref{sec:segment_planner}--\ref{sec:lane_binder}).

\item We describe software acknowledgments and coordinated link
configuration and recovery for executing lane-level connections, and
prototype software acknowledgments and verify configuration changes
on commodity server NICs and an OCS
(\S\ref{sec:endpoint_execution}).

\item We evaluate scheduling gains, module contributions, and
multi-OCS mapping in simulation, together with communication replay,
training, and software-ACK performance on a hardware testbed
(\S\ref{sec:evaluation}).

\end{itemize}

\section{Background and Motivation}
\label{background-and-motivation}

\begin{figure*}[t]
    \centering
    \begin{minipage}[t]{0.49\textwidth}
        \centering
        \includegraphics[width=\linewidth]
            {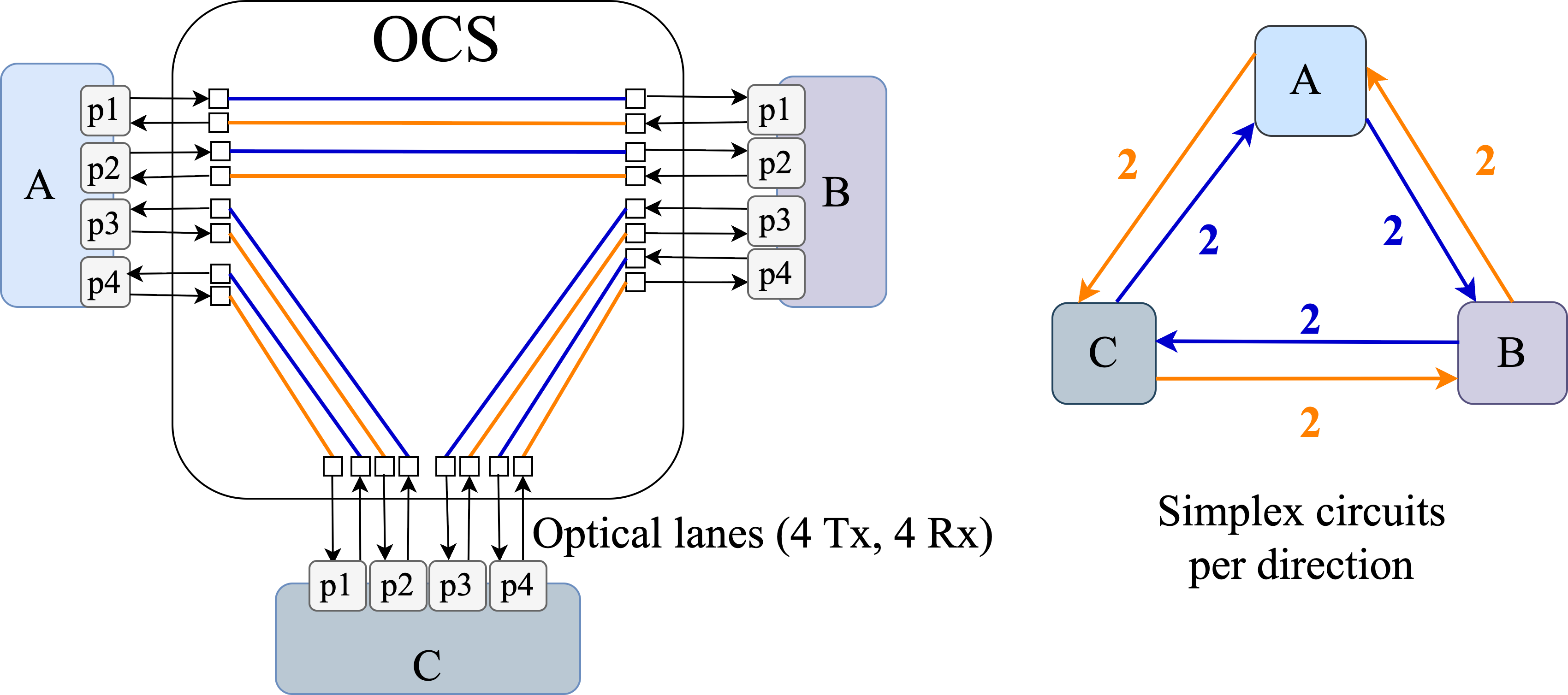}
        \par\smallskip
        {\small (a) Duplex scheduler}
    \end{minipage}
    \hfill
    \begin{minipage}[t]{0.49\textwidth}
        \centering
        \includegraphics[width=\linewidth]
            {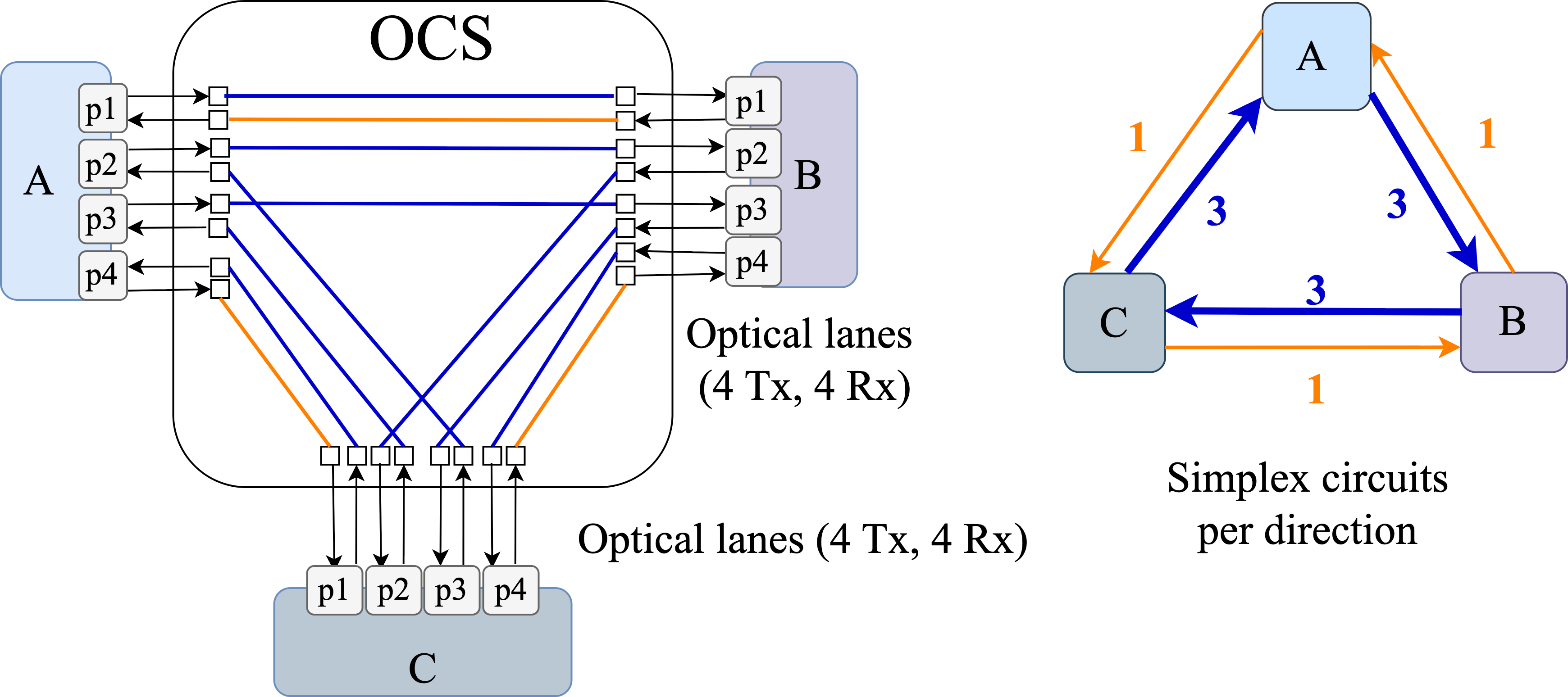}
        \par\smallskip
        {\small (b) Simplex scheduler}
    \end{minipage}
    \caption{Duplex and simplex scheduling with the same four TX
    lanes and four RX lanes per node. Each panel shows the OCS
    connections (left) and the resulting directed logical links
    (right). Duplex matching assigns two simplex circuits to each
    direction of every node pair. Lane-level matching assigns
    three along \(A\!\rightarrow\!B\!\rightarrow\!C\!\rightarrow\!A\)
    and one along the reverse cycle.}
    \Description{Two diagrams compare duplex and simplex scheduling
    among three nodes. Duplex scheduling produces a 2:2 circuit
    allocation, while simplex scheduling produces a 3:1 allocation
    using the same TX and RX lane inventories.}
    \label{fig:duplex_simplex}
\end{figure*}

\subsection{OCS Fabrics and Terminology}
\label{nodes-ports-and-optical-lanes}
\textbf{OCS hardware.} An optical circuit switch (OCS) is a layer-0 device: it forwards light, not packets. Commercial 3D-MEMS OCSs provide \textit{P} ports in a non-blocking crossbar; an array of micro-mirrors steers the beam arriving at any input to any output, establishing a transparent, full-bandwidth light path agnostic to the modulation and bit rate carried on it~\cite{polatis7000, calient_s320}.
Two hardware facts matter for this paper. First, every cross-connect is
physically \textit{simplex}: it carries light in exactly one direction, and what
is normally called a bidirectional "circuit" is in fact two independent
cross-connects installed by two coordinated mirror settings.
Second, reconfiguring the mirror array takes on the order of tens of
milliseconds~\cite{farrington2011hardware,dickson2004mems,mori2021highport}, so the fabric operates in epochs: a scheduler computes a topology, the switch installs it, and the topology remains fixed until the next epoch.

\textbf{Ports and lanes.} Each OCS port terminates a fiber pair—one fiber carries light into the switch, the other carries light out—attached to a duplex transceiver at the endpoint. A port therefore bundles exactly two simplex resources: a transmit (TX) \textit{lane} and a receive (RX) \textit{lane}, and a \textit{P}-port switch exposes \textit{2P} independently steerable lanes. (Throughout this paper, "lane" denotes one direction of a port's fiber pair, not a SerDes or wavelength lane inside a transceiver.) We call a light path from one TX lane to one RX lane a \textit{simplex circuit}, and the conventional pair of simplex circuits between two ports a \textit{duplex circuit}.

\textbf{Nodes and fabric boundary.}
A node is either an OCS-facing EPS, aggregating its attached servers'
traffic~\cite{Apollo,JupiterEvolving}, or a directly attached
server~\cite{topoopt,mixnet,actina}. Its \(k_n\) OCS-facing ports provide \(k_n\) TX
and \(k_n\) RX lanes. The OCS fabric comprises the optical switches
and connecting fibers, with actual wiring and path constraints retained
in \S\ref{sec:lane_binder}. Figure~\ref{fig:node_fabric_mapping} in Appendix~\ref{app:eps_sensitivity} illustrates the two deployment mappings. These deployment mappings do not imply
independent TX/RX hardware support (\S\ref{sec:discussion}).

\textbf{Scheduling model and the duplex convention.} Demand-aware OCS scheduling proceeds epoch by epoch. The scheduler takes as input a directed, node-level demand matrix $D^{(k)}= [d_{ij}]$, where $d_{ij}^k$ is the byte volume at which node \textit{i} wishes to send to node \textit{j} in communication operation \(k\), and outputs a topology subject to port-count and reconfiguration constraints. All demand-aware OCS fabrics we are aware of express the topology as a \textit{duplex matching}: an undirected matching over ports in which connecting port \textit{p} to port \textit{q} installs both the $p\!\rightarrow\!q$ and $q\!\rightarrow\!p$ cross-connects atomically~\cite{actina,mixnet,infinitehbd,opus,lumoscore, tpuv4,topoopt}. Lane-granularity scheduling (§3) instead computes two \textit{directed matchings}—one from TX lanes to RX lanes in each direction—while preserving \textit{node-level duplex}: every node keeps at least one active TX lane and one active RX lane. Table~\ref{tab:terminology} summarizes the terminology and notation used throughout the paper.

\begin{table}[!t]
\caption{Optical fabric terminology and notation.}
\label{tab:terminology}
\small
\begin{tabular}{@{}p{0.29\columnwidth}p{0.67\columnwidth}@{}}
\toprule
Term / symbol & Meaning \\
\midrule
Node \(n\), \(k_n\) & Server or EPS with \(k_n\) attached duplex ports. \\
Port; \(P\) & Duplex endpoint interface; total attached port count. \\
TX / RX lane; \(c\) & Fixed transmit / receive resource; lane rate. \\
Simplex circuit & One TX-to-RX light path in use. \\
Duplex circuit & Reciprocal pair of simplex circuits between ports. \\
\(X_{uv}\) & Number of circuits on directed logical link \(u\to v\). \\
\(D^{(k)}\) & Directed node-pair byte demands of operation \(k\). \\
Configuration & Set of simultaneous TX-to-RX lane bindings. \\
Segment & Fixed-configuration interval / consecutive operations assigned to it. \\
Reconfiguration delay & Exposed time to change configuration and restore usable communication. \\
\bottomrule
\end{tabular}
\end{table}

\begin{figure*}[t]
    \centering
    \includegraphics[width=\textwidth]{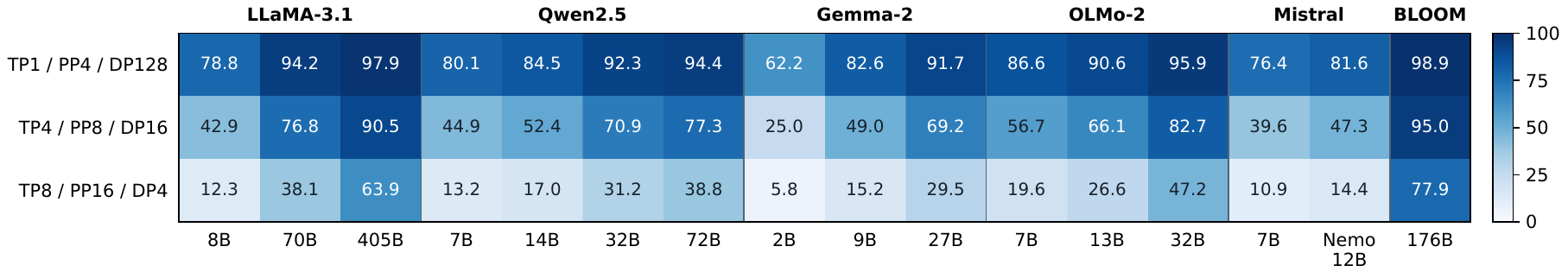}

    \caption{Traffic share on strongly skewed
    ($\rho_{uv}>4$) node pairs across 48 complete mixed
    schedules at world size 512.}
    \Description{A heatmap compares 16 models across three TP, PP, and DP configurations
  at world size 512. Cells show the fraction of traffic carried by
  strongly skewed fabric-node pairs; darker blue indicates a larger
  fraction. Shares range from 5.8 to 98.9 percent, with a median of 65.0
  percent.}
    \label{fig:bg_cross_model_skew}
\end{figure*}

\subsection{Directional Demand Exists Between Nodes}
\label{directional-demand-exists-at-the-node-boundary}

Large-model training distributes computation across ranks, i.e., the
participants in a parallel execution. Tensor parallelism (TP) partitions
tensor computations across ranks; data parallelism (DP) maintains model
replicas and exchanges their updates through collectives such as
AllReduce, ReduceScatter, and AllGather; pipeline parallelism (PP)
places successive model stages on different ranks and exchanges
activations and gradients between them. In the deployments considered in
this paper, each TP group fits entirely within one server. TP
communication therefore never traverses the
optical fabric, whether a node is a server or an EPS. The inter-node
demand considered here comes from the DP and PP transfers.

One collective communication seems to be symmetric: each node
may transmit and receive equal total volumes. However, when viewed
as node pairs, the same exchange can be asymmetric. In a selected ring
\(A\rightarrow B\rightarrow C\rightarrow A\), A sends bulk data to B but
receives bulk data from C. Equal total sends and receives at A do not
require equal demand on \(A\!\rightarrow\!B\) and \(B\!\rightarrow\!A\).

Figure~\ref{fig:bg_cross_model_skew} examines how much training traffic exhibits this pairwise imbalance. We analyze 48 graph-derived mixed training schedules across 16 dense models~\cite{llama3,qwen25,gemma2,olmo2,olmo2_32b,mistral7b,mistral_nemo,bloom}, each using 512 GPUs organized into 64 eight-GPU
servers. For each schedule, we aggregate transfers from all ranks by
ordered node pair, exclude intra-node traffic, and include the
calibrated forward and reverse control overheads. Let \(D_{uv}\) and
\(D_{vu}\) denote the resulting byte volumes in the two directions over
the complete schedule. We identify a pair as \textbf{strongly skewed} when

$\rho_{uv}=\max\{D_{uv},D_{vu}\}/{\min\{D_{uv},D_{vu}\}}>4$.
Each heatmap cell reports the fraction of
inter-server bytes carried by strongly skewed pairs, counting both
directions. It ranges from 5.8\% to 98.9\%, with a median of
65.0\%. Thus, in the evaluated schedules, much of the traffic crosses
node pairs whose directional demands differ substantially.

For switch nodes, the same calculation aggregates across all servers
attached to each switch. Aggregation can absorb transfers within a switch
or combine traffic in opposite directions, so server-level skew does not
automatically carry over to switch pairs. Grouping four consecutive servers
per switch in the same workloads gives a median strongly skewed traffic
share of 47.7\%; grouping eight gives 0\% over the complete schedules.
The latter does not imply exact symmetry or rule out skew within
individual phases.
\begin{figure}[t]
    \centering
    \includegraphics[width=0.9\linewidth]
        {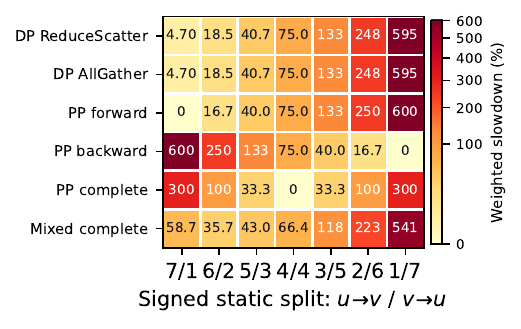}

    \caption{Traffic-weighted slowdown of fixed directional capacity splits relative to the best pair-local split in each communication window. This analysis isolates directional allocation and does not enforce shared node-level lane inventories}
    \Description{A heatmap compares seven fixed directional splits
    across six communication windows. DP and forward PP favor 7:1,
    backward PP favors 1:7, and complete PP favors 4:4. No single
    split is best across all windows.}
    \label{fig:bg_fixed_split}
\end{figure}

\subsection{OCS Hardware Supports Lane-Level Matching}
\label{OCS-support-lane-matching}

The directional demand above can be served by assigning different numbers
of simplex circuits to opposite directions of a node pair. Under duplex
matching, connecting port \(p\) at node \(u\) to port \(q\) at node \(v\)
reserves both \(p_{\mathrm{TX}}\!\rightarrow q_{\mathrm{RX}}\) and
\(q_{\mathrm{TX}}\!\rightarrow p_{\mathrm{RX}}\). With equal-rate lanes,
this enforces \(X_{uv}=X_{vu}\), coupling the capacity of the two directions
even when their demands differ.

The separate TX and RX lanes provide a way to remove this coupling.
When these lanes attach to independently configurable OCS interfaces,
the switch can connect a port's TX lane to an RX lane at one node while
connecting its RX lane from a TX lane at another
node~\cite{polatis6000,polatis7000,calient_s320}.
For example, a port at node A can transmit to B and receive from C.
Across its ports, A can therefore allocate more TX lanes to B and more
RX lanes to C, while maintaining node-level duplex. This capability
applies to both server nodes and switch nodes: the OCS configures optical
paths between their lanes, regardless of whether those lanes terminate
on NICs or EPS uplinks. Each node retains its fixed inventory of
\(k_n\) TX lanes and \(k_n\) RX lanes; lane-level matching changes their
peers, not their transmit or receive roles.

Consider three nodes, each with four ports. Suppose the directions
\(A\!\rightarrow\!B\), \(B\!\rightarrow\!C\), and
\(C\!\rightarrow\!A\) each carry \(3V\) bytes, while each reverse
direction carries \(V\). A duplex matching can allocate two simplex
circuits to each direction of every node pair. Lane-level matching can
instead allocate three along the forward cycle and one along the reverse
cycle. Both configurations use four TX lanes and four RX lanes
at every node. At a per-lane rate of \(c\) bytes per second, assuming
concurrent transfers with no other bottleneck, the transfer time falls
from \(3V/(2c)\) to \(V/c\). The improvement comes from matching the same
optical resources to the directional demand.

Optical support alone does not ensure usable connections. Helios explored unidirectional circuits~\cite{farrington2010helios}, and measurements show that simplex reconfiguration can trigger Ethernet link-failure handling~\cite{lane_zerwas}. NICs and EPS uplinks must remain operational with different TX and RX peers, and feedback must reach the sender. Exploiting this capability therefore requires link and feedback support alongside scheduling.

\begin{figure*}[t]
    \centering
    \includegraphics[width=0.94\linewidth]{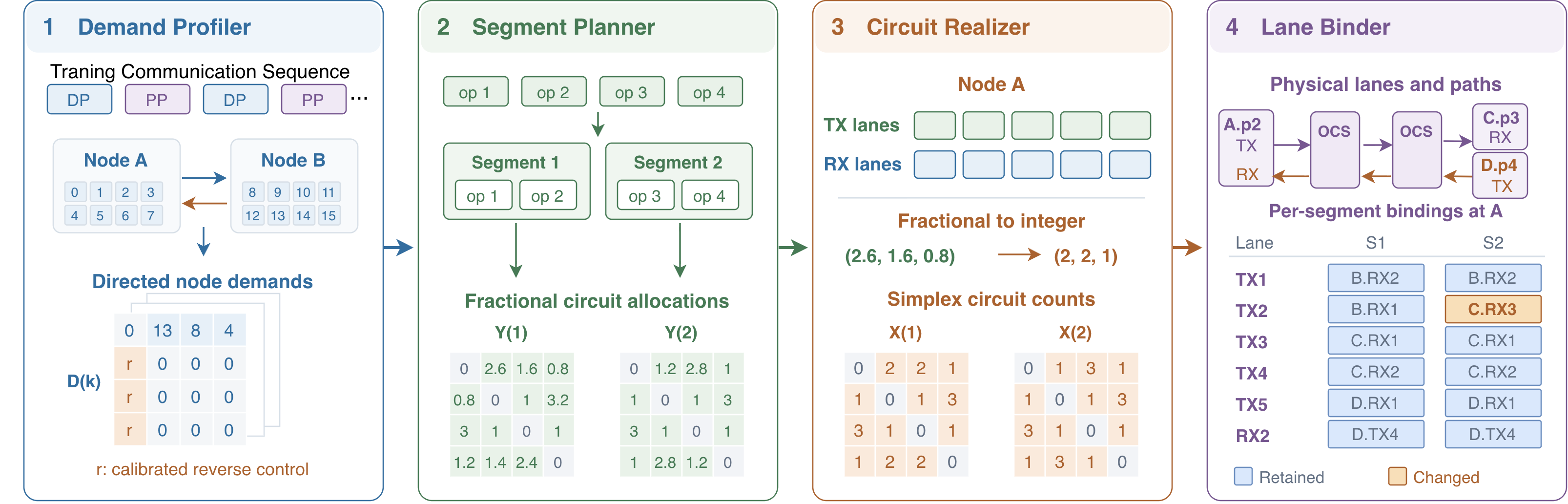}
     \caption{LACE converts a training communication schedule into
    physical lane connections through four stages.}
    \Description{The Demand Profiler produces directed demand matrices;
    the Segment Planner groups operations and computes fractional
    allocations; the Circuit Realizer produces integer counts; the
    Lane Binder assigns physical lanes and optical paths.}
    \label{fig:design_overview}
\end{figure*}

\subsection{Asymmetric Connectivity Needs Scheduling}
\label{why-asymmetric-connectivity-needs-scheduling}

The hardware capability above raises a scheduling question: how should
lane allocations follow demand? DP collectives and PP transfers change
node-pair traffic in both volume and direction, so an allocation that
fits one communication window may not fit the next.

Figure~\ref{fig:bg_fixed_split} shows why one fixed allocation
is not enough. For each communicating pair, we compare seven
allocations of eight simplex circuits between the two directions,
from seven in one direction and one in the other (\(7:1\)) to
the reverse (\(1:7\)). We compare the transfer time under each
allocation with that under the best allocation for the same
pair and communication window, and average the slowdown weighted
by traffic volume.
DP ReduceScatter, DP AllGather, and forward PP perform best with
\(7:1\), while backward PP performs best with \(1:7\), and complete
PP with \(4:4\). No single fixed allocation performs best across all the evaluated
communication windows, motivating lane allocations that adapt
during training.

One possible way to avoid the limitations of fixed allocations
is to make communication demand symmetric. Existing collective
optimizations, however, do not guarantee this.
NCCL selects algorithms and channels for efficient
communication~\cite{nccl,demystifyingNCCL}, but equal total sends
and receives at a node do not imply equal traffic in both
directions of each node pair.
Bidirectional Ring AllReduce and RingBiOdd can balance payload
across opposite-direction rings, but require suitable connectivity
and collective-specific schedules~\cite{ringbiodd}; this balance
does not extend automatically to other communication operations.
SCCL and TACCL synthesize topology-aware collective
schedules~\cite{sccl,taccl}, but minimizing communication time
does not require pairwise symmetric traffic.
AdapCC adapts communication to resource and network
changes~\cite{adapcc}, while ResCCL improves resource use through
task scheduling~\cite{resccl}; neither objective guarantees equal
opposite-direction traffic between nodes.
Thus, even when individual collectives are optimized or balanced,
asymmetric demand can remain as DP collectives and PP transfers
execute and overlap.

Since asymmetric demand remains, lane connections must be chosen
jointly across node pairs and time. Each simplex circuit consumes a
source TX lane and a destination RX lane, so the pairwise choices in
Figure~\ref{fig:bg_fixed_split} must fit shared node inventories.
Changing connections also incurs optical reconfiguration and node
recovery delay. Consecutive operations should therefore share a
configuration unless changing it saves more transfer time than it costs.
The next section presents LACE, which jointly chooses segment boundaries
and lane allocations to minimize communication completion time,
including reconfiguration delay.

\providecommand{\sysname}{LACE}
\section{\sysname{} Design}
\label{sec:design}

\subsection{From Communication Operations to Lane Connections}
\sysname{} takes an ordered communication schedule, collective algorithms,
rank placement, node port inventories, and the physical OCS topology.
It outputs which operations share a configuration and the lane
connections used by each configuration.
Figure~\ref{fig:design_overview} shows four stages.
The \textbf{Demand Profiler} aggregates transfers into node-level demand.
The \textbf{Segment Planner} groups consecutive operations and computes
fractional circuit allocations, balancing transfer and reconfiguration
costs. The \textbf{Circuit Realizer} converts them into integer counts,
and the \textbf{Lane Binder} assigns physical lanes and optical paths.
Nodes execute these connections using \textbf{software acknowledgments}
and coordinated \textbf{link configuration and recovery}.

\subsection{Constructing Node-Level Demand}
The profiler expands collectives into directed rank transfers and maps
ranks to server or switch nodes. For each operation, it sums bytes by
ordered node pair to form \(D^{(k)}\), excluding transfers within a node
and combining transfers across its ports. It preserves operation order
and keeps opposite directions separate. Calibrated reverse control
traffic is included to reserve return connections for feedback
(\S\ref{sec:endpoint_execution}).
Appendix~\ref{app:planning_formulation} formalizes the demand and
resource constraints.

\subsection{Planning Segments and Fractional Allocations}
\label{sec:segment_planner}
The Segment Planner uses fractional allocations to compare candidate
segments without solving an integer problem for every candidate.
These targets estimate the cost of sharing a configuration; the
Circuit Realizer makes them executable afterward. Starting with one
segment per operation, the planner merges adjacent segments when
sharing costs less than the reconfiguration it avoids.

\textbf{Evaluating a shared allocation.}
For segment \(I\), let \(Y_{uv}\) be the fractional number of
simplex circuits from \(u\) to \(v\). Each allocation must fit
within the source's TX lanes and the destination's RX lanes,
with positive allocations for all demanded directions.
At per-lane rate \(c\), the estimated segment time is
\begin{equation}
C(I,Y)=\sum_{k\in I}
\max_{d_{uv}^{(k)}>0}\frac{d_{uv}^{(k)}}{cY_{uv}}.
\label{eq:segment_time}
\end{equation}
Transfers within an operation proceed concurrently, so its slowest
transfer determines completion. Summing these times models operations
executing sequentially in the selected schedule.

Consider demands \((8,2)\) and \((6,2)\) on
\((A\!\rightarrow\!B,A\!\rightarrow\!C)\), with four TX lanes at A,
sufficient RX lanes at B and C, and \(c=1\). Separate allocations
\((3.2,0.8)\) and \((3,1)\) finish in \(2.5\) and \(2\) time units.
Sharing \((3,1)\) takes \(8/3+2\approx4.67\), slightly longer than
\(4.5\), but avoids a reconfiguration.

\textbf{Finding a fractional allocation.}
The allocator normalizes each operation's demands by its largest
entry, sums them by direction, and uses the square roots as initial
weights. Larger weights receive more fractional circuits, subject to
both lane inventories. It then evaluates the original byte volumes
and reweights the slowest or nearly slowest transfers. After a fixed
number of rounds, it returns the best allocation found and its
estimated time, \(\widehat C(I)\).

\textbf{Merging adjacent segments.}
For adjacent segments \(I,J\), the saving is
\begin{equation}
\operatorname{save}(I,J)=
\widehat C(I)+\widehat C(J)+\delta-\widehat C(I\cup J),
\label{eq:merge_saving}
\end{equation}
where \(\delta\) is the reconfiguration delay. In the example,
\(\delta=0.5\) makes merging save \(2.5+2+0.5-4.67\approx0.33\).
If the next operations have demands \((2,8)\) and \((2,6)\), they
similarly benefit from sharing \((1,3)\). This produces two segments,
\([1,2]\mid[3,4]\), taking \(4.67+0.5+4.67\approx9.83\) time units.
Combining all four would take \(14\) under their best shared allocation
\((2,2)\), so the planner retains the two segments. A priority queue selects the largest positive saving. After each merge,
only candidates involving the new segment and its neighbors are
recomputed; planning stops when no saving remains.
Appendix~\ref{app:segment_dp} details the fractional allocator
and segment-merging procedure.

\subsection{Realizing Integer Simplex Circuits}
\label{sec:ocs_realizer}

The Circuit Realizer converts fractional allocations \(Y\) into integer
counts \(X\), seeking communication time within \(1+\epsilon\) of the
planner's estimate under the node lane inventories. Here, \(\epsilon\)
is the allowed increase in communication time.

\textbf{Why direct rounding is insufficient.}
Suppose A sends \((13,8,4)\) units to B, C, and D, with five TX lanes,
sufficient destination RX lanes, and \(c=1\). The fractional allocation
\((2.6,1.6,0.8)\) finishes in \(5\) time units, but rounding gives
\((3,2,1)\), requiring six lanes. Removing one A-to-B circuit gives
\((2,2,1)\) and time \(6.5\); removing one A-to-C circuit instead gives
\((3,1,1)\) and time \(8\). Removing A-to-D leaves demand unserved. Thus, repairing rounded allocations requires preserving demand coverage and evaluating transfer time, rather than merely satisfying lane budgets.

\textbf{Constructing and pruning an allocation.}
The realizer first assigns one circuit to every demanded direction.
If this exceeds an inventory, the segment cannot be served directly
by one configuration. Otherwise, it adds circuits in order of largest
positive gap \(Y_{uv}-X_{uv}\), with a stable node-pair order for ties,
checking source TX and destination RX availability. If the tolerance
remains unmet and lanes are available, it attempts additions with the
largest reduction in segment time.

Once the tolerance is met, the realizer tries removing circuits in
order of smallest time increase, accepting only removals that preserve
coverage and the tolerance. For the example, \((2,2,1)\) meets
\(\epsilon=30\%\); any removal would violate coverage or this tolerance.
If the search cannot meet the tolerance, it reports the shortfall
with the best feasible allocation found. It passes integer counts to
the Lane Binder. Appendix~\ref{app:discrete_realization} gives the
constraints and time changes used to evaluate additions and removals.

\subsection{Binding Circuits Across the OCS Fabric}
\label{sec:lane_binder}

The Lane Binder assigns each circuit a source TX lane, a destination
RX lane, and an optical path. With one nonblocking OCS, it pairs
\(X_{uv}\) distinct TX lanes at \(u\) with RX lanes at \(v\), using each
lane at most once. Multiple OCSes add internal path constraints.

\textbf{Routing and repair through multiple OCSes.}
The binder expands circuit counts into individual requests and
jointly selects their lanes and paths. Each path reserves its TX/RX
lanes, inter-switch fibers, and OCS cross-connects; no two circuits
may occupy the same directional resource. Two requests can therefore
conflict on an internal fiber even when both nodes have free lanes.
The binder first reroutes circuits, including previously assigned
paths, while preserving the requested counts.

For the evaluated three-stage Clos, the binder represents
inter-leaf circuit requests as a bipartite multigraph:
source leaves form one side, destination leaves the other,
and each requested circuit contributes an edge.
An edge color selects a middle plane, with edges sharing
a source or destination leaf assigned different colors.
This constructs conflict-free paths when enough middle
planes are available.

When internal resources are constrained, the binder first
attempts rerouting, including previously assigned paths,
while preserving the requested counts.
If routing still fails, it removes an optional circuit with the
smallest increase in segment time and retries. A removal must leave
every demanded direction served. The binder checks the resulting
time against the Realizer's tolerance and reports any excess.
If minimum coverage cannot be routed, it returns the segment for
replanning. A heuristic routing failure is an unresolved mapping,
not proof of physical infeasibility.

\textbf{Completing connections at active ports.}
The targeted commodity nodes require both lanes of each enabled port
to be connected, potentially to different peers. The binder completes
any unconnected lane using available resources. These added circuits
need not carry data, but consume lanes and optical paths and are
checked together with data-carrying circuits. Unused ports remain
disabled. If completion fails, the binder revises lane assignments
or returns the configuration for reallocation. Node readiness is
verified before transmission (\S\ref{sec:endpoint_execution}).

\textbf{Retaining connections across segments.}

The binder prefers existing bindings and paths. For example, changing
\((2,2,1)\) to \((1,3,1)\) can retain four circuits and redirect only
one A-to-B circuit toward C. In a multi-OCS fabric, retaining a circuit
also requires retaining its path; paths may move when they block new
requests. The output specifies lane bindings and cross-connects for
each OCS. Appendix~\ref{app:constrained_routing} details port-completion
accounting, the Clos path construction, and reuse and repair conditions.

\subsection{Executing Lane-Level Configurations}
\label{sec:endpoint_execution}

Executing lane bindings requires nodes to confirm
delivery through potentially different ports and keep
ports operational when their TX and RX lanes connect
to different nodes.

\textbf{Software acknowledgments.}
For server nodes, \sysname{} uses RDMA Unreliable Connected (UC)
transfers with software acknowledgments. UC retains NIC data movement
while placing reliability in software, allowing feedback received by
one NIC to confirm data sent by another without sharing hardware
queue-pair state. UC is an implementation choice, not a requirement
of lane-level scheduling. UCCL similarly uses UC with software
reliability for GPU collective communication~\cite{uccl}.

Chunks carry transfer IDs and sequence numbers. The receiver records
completed chunks and batches cumulative acknowledgments, flushing on
a timer or transfer completion. Each NIC owns its queue pairs and
registered buffers; host software tracks completion across NICs.

Calibrated reverse demand makes the Realizer reserve at least one
reverse simplex circuit, so feedback returns directly between
communicating nodes, potentially through a different port. An EPS
forwards feedback using configured routes. When direct allocation is
infeasible, forwarding through intermediate nodes is a fallback,
subject to available paths and resources.

The sender bounds outstanding data and retransmits after timeout;
the receiver suppresses duplicates. A transfer completes only after
all chunks are confirmed; source buffers are released only after
local send completion and acknowledgment.
\S\ref{sec:eval-testbed} measures software-ACK performance on a single
100~Gbps port; multi-NIC runtime integration remains future work.
\S\ref{sec:discussion} discusses compatibility with RC.
Separate return paths have precedent in RFC~3077's tunneling over a
bidirectional network~\cite{rfc3077}; \sysname{} instead uses reverse
simplex circuits within the OCS fabric.

\textbf{Link configuration and recovery.}
Simplex reconfiguration can disrupt Ethernet link-failure
handling~\cite{lane_zerwas}. Disabling autonegotiation addresses
parameter negotiation, but not dependence on a working receive signal.
The controller, which selects both ends of every circuit, provisions
compatible rate and FEC settings from NIC, EPS, and module capabilities.
It disables autonegotiation only on ports verified to support fixed
settings and retains required link training. The Lane Binder supplies
the active-port connections described in \S\ref{sec:lane_binder}.

Reconfiguration can still interrupt receive signals. Nodes drain
outstanding transfers over the old connections before the controller
installs new circuits and forwarding rules. It checks port status and
sends configuration-ID probes; receiving nodes report through the
management network. Transmission resumes only after verification.
On timeout, the controller attempts bounded port reinitialization,
then restoration of the previous configuration if necessary, keeping
transmission paused until verification succeeds. Hardware fault
detection remains enabled. The Segment Planner charges the exposed
draining, switching, and recovery time as reconfiguration delay.

\section{Evaluation}
\label{sec:evaluation}

We evaluate lane-level scheduling through fixed-topology replay and
training on a three-server OCS testbed, separate software-ACK and
recovery experiments, and large-scale simulations of graph-derived
LLaMA-3.1 workloads~\cite{llama3}.

\subsection{Physical Testbed Evaluation}
\label{sec:eval-testbed}

\textbf{Setup.}
Three GPU servers connect to an $8\times8$ Polatis OCS, each through
two 100~Gbps NIC ports with separate TX and RX fibers. We compare LACE with ACTINA~\cite{actina}, a state-of-the-art
OCS scheduling approach for distributed AI training. ACTINA and LACE
compute their topologies offline: a symmetric triangle and a directed
double ring, respectively. Each remains fixed throughout replay or training.
Both methods use TCP with the same striping policy.
Appendix~\ref{app:testbed} gives the workload and transport details.

\begin{figure}[t]
  \centering
  \includegraphics[width=\columnwidth]{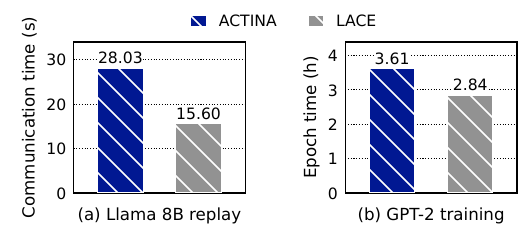}

  \caption{Fixed-topology testbed performance.
  (a) Median complete replay time over two runs.
  (b) One GPT-2 training epoch per topology.}
  \Description{Replay time decreases from 28.03 to 15.60 seconds;
  training time decreases from 3.61 to 2.84 hours.}
  \label{fig:eval-testbed-performance}
\end{figure}

\textbf{Replay and training.}
We replay the complete mixed DP/PP schedule of a LLaMA-3.1 8B workload
with TP=4, PP=2, DP=3 and eight microbatches. Its 67 slots represent 24 logical GPUs. LACE achieves a $1.80\times$
replay speedup (Figure~\ref{fig:eval-testbed-performance}). Although bidirectional 1F1B PP is 49.9\% slower,
the DP improvement reduces total replay time by 44.3\%.
We also fine-tune GPT-2 Small on WikiText-103 with one GPU per server,
identical initial weights and data order.
Over 4,846 optimizer steps, training speeds up by $1.27\times$. Held-out losses are 2.8169 and 2.8166,
with no skipped updates.

\begin{figure}[t]
  \centering
  \includegraphics[width=\columnwidth]{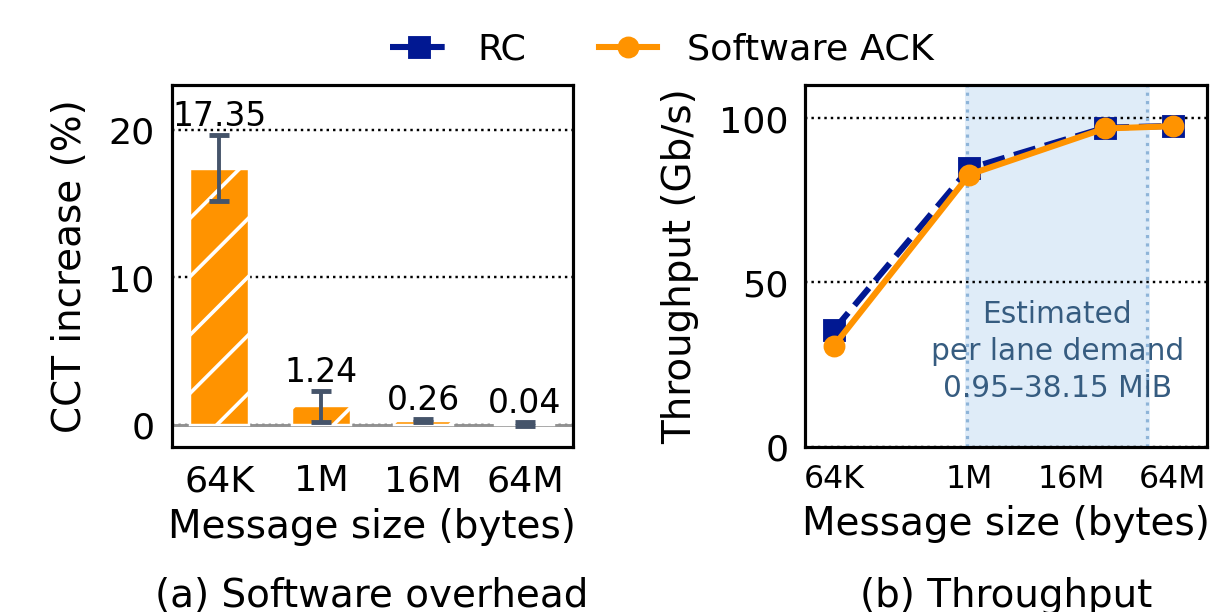}
  \caption{Software ACKs: 30 paired trials per size.
  (a) Mean CCT increase over RC with 95\% bootstrap intervals.
  (b) Median throughput; shading marks estimated per-lane payloads
  (Appendix~\ref{app:ack_payload_estimate}). K/M denote KiB/MiB.}
  \Description{Software-ACK overhead falls from 17.35 percent at 64 KiB
  to 0.04 percent at 64 MiB. Both methods approach 100 Gbps on large messages.}
  \label{fig:eval-software-ack}
\end{figure}

\begin{figure*}[t]
  \centering
  \includegraphics[width=\textwidth]{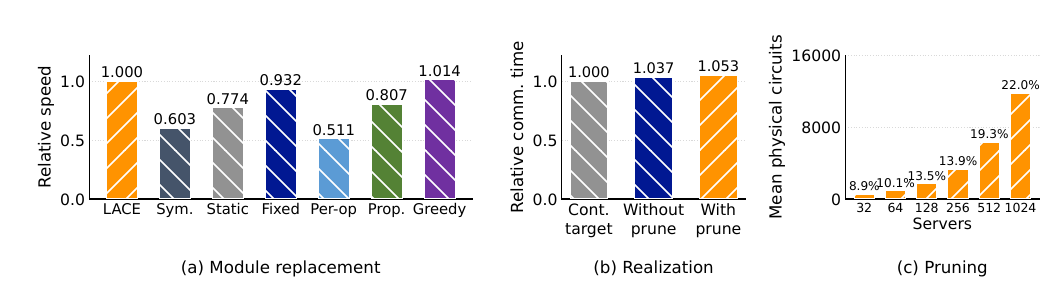}

  \caption{Module ablation (70B Full, 128 servers).
  (a) Speed normalized to LACE. (b) Communication time normalized to
  the fractional target. (c) Speed and mean physical circuits
  normalized to no pruning.}
  \Description{Module replacements reduce speed except Greedy, which
  omits pruning. Integer communication time is 3.72 percent above the
  fractional target without pruning and 5.29 percent above with pruning.
  Pruning retains 98.64 percent speed and 86.51 percent physical circuits.}
  \label{fig:eval-validation}
\end{figure*}

\textbf{Software acknowledgments.}
We compare RDMA RC with UC plus cumulative software acknowledgments using the same port for data and feedback.
Both use host buffers and one busy-polling thread per host, with
no additional ACK thread (Figure~\ref{fig:eval-software-ack}). At 64 MiB, both approach 98~Gbps. The median combined endpoint CPU cost is 0.1759 CPU-s/GiB for RC and 0.1760 for UC with software ACKs. These process-level costs include polling and do not isolate ACK-processing cycles (Appendix~\ref{app:testbed}).

Under the Megatron configuration in Appendix~\ref{app:ack_payload_estimate}, estimated per-lane DP/PP payloads are 0.95--38.15 MiB, for which our 100~Gbps measurements show modest overhead.

\subsection{Simulation Methodology}
\label{sec:eval-methodology}

\textbf{Workloads and node resources.}
The main experiments use LLaMA-3.1 70B and 405B Full schedules on
32--512 eight-GPU servers. TP=8 stays within each server, so node-level
demand contains only inter-server DP and PP transfers. The models use
PP=8 and PP=16, respectively, with DP set by server count. The Demand
Profiler preserves collective execution, rank placement, and operation
order, and aggregates each operation's rank transfers by server pair.
All methods receive the same calibrated forward and reverse control
bytes from NCCL/RoCE measurements of our testbed.

Each server node has 16 ports: 16 TX lanes and 16 RX lanes, each at
400~Gbps. A sensitivity experiment uses eight 800-Gb/s ports, preserving
6.4~Tbps per direction. Defaults are $\delta=20$ ms and $\epsilon=0.05$.
The Lane Binder completes both directions of every enabled port and
checks lane inventories; additional connections consume resources but
do not contribute extra data throughput.

\textbf{Methods and metric.}
We compare LACE with the duplex OCS scheduler ACTINA~\cite{actina},
using identical demand, ports, line rates, and reconfiguration delay.
Symmetric LACE couples opposite-direction circuit counts at LACE's
segment boundaries, isolating directionality without independently
optimizing a symmetric schedule. An ideal nonblocking bound, limited only by each
node's aggregate TX/RX rates, provides context.

CCT sums operation completion times using the integer simplex circuit
counts in Equation~\ref{eq:segment_time}, plus $\delta$ between different
configurations. Transfers within an operation proceed concurrently;
operations follow the selected order. Identical adjacent configurations
merge before charging reconfiguration. The model excludes computation,
compute/communication overlap, packet dynamics, and host-side transport processing limits, including software-ACK service capacity. Simulated CCT gains
therefore measure communication scheduling, not training speedup.

\begin{figure}[t]
  \centering
  \includegraphics[width=\columnwidth]{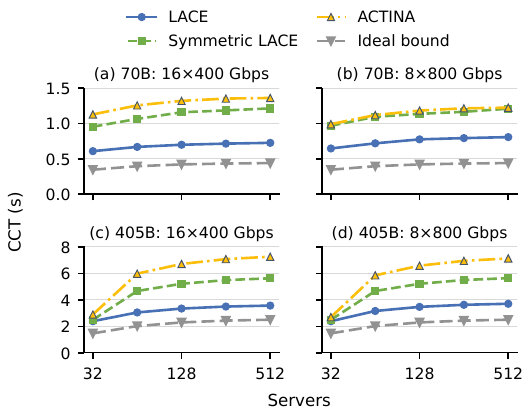}
  \caption{Full-schedule CCT at equal aggregate bandwidth:
  6.4~Tbps per server per direction.}
  \Description{Four line charts compare LACE, Symmetric LACE, ACTINA,
  and an ideal bound over 32 to 512 servers for 70B and 405B.
  LACE has lower CCT than both optical comparisons at all plotted points.}
  \label{fig:eval-cct}
\end{figure}

\subsection{Communication Completion Time}
\label{sec:eval-cct}
\textbf{Lane-level scheduling improves directional communication.}
At 16 ports per server, LACE speeds up ACTINA by $1.86$--$1.89\times$
for 70B and $1.21$--$2.04\times$ for 405B
(Figure~\ref{fig:eval-cct}). Relative to the same-boundary symmetric
ablation, gains are $1.57$--$1.67\times$ for 70B. For 405B, the gain
is only $1.04\times$ at 32 servers, where DP=2 makes DP transfers
symmetric; it reaches $1.53$--$1.58\times$ at 64--512 servers.
Thus the benefit depends on the directional demand remaining in the
selected schedule. At 512 servers, LACE completes the 70B and 405B
schedules in 0.726 and 3.569 s, respectively.

\textbf{More ports improve allocation at the same total rate.}
As shown in Figure~\ref{fig:eval-cct}, using eight 800~Gbps ports instead of sixteen 400~Gbps ports raises LACE CCT
by 6.33--11.19\% for 70B and 3.66--3.83\% for 405B at 64--512 servers.
More ports permit smaller circuit-allocation steps and more simultaneous
peers. The 405B, 32-server case is an exception: eight ports are
0.05\% faster, reflecting the heuristic integer allocation rather than
a strictly monotonic guarantee.

\subsection{Module Ablation}
\label{sec:eval-validation}

\textbf{Directionality, segmentation, and allocation each matter.}
Figure~\ref{fig:eval-validation} (a) uses 70B Full at 128 servers.
Coupling opposite directions retains 60.27\% of LACE's speed.
Replacing adaptive segmentation with one static configuration, the
best fixed-length partition, or one configuration per operation reduces
speed by 22.57\%, 6.83\%, and 48.88\%, respectively. The best fixed
length is 50 among $\{2,5,10,20,50,100\}$. Allocating fractional circuits
in proportion to aggregate demand within the same segments reduces
speed by 19.27\%. These controlled replacements show why both changing
demands and the cost of changing configurations matter.

\textbf{Pruning saves circuits with a small time penalty.}
At 128 servers, communication time exceeds the fractional target
by 3.72\% without pruning and 5.29\% with it
(Figure~\ref{fig:eval-validation}(b)).
Pruning reduces mean physical simplex circuits per configuration
from 1,928 to 1,668 (13.49\%), including port completion,
at a 1.36\% speed loss.
Circuit savings rise from 8.9\% at 32 servers to 22.0\% at 1024
servers (Figure~\ref{fig:eval-validation}(c)).
Greedy matches LACE under the same pruning setting, demonstrating
the pruning tradeoff rather than an additional gain from
performance augmentation.

\subsection{Deployment on Optical Fabrics}
\label{sec:eval-multi-ocs}

\begin{figure}[tbp]
  \centering
  \includegraphics[width=\columnwidth]{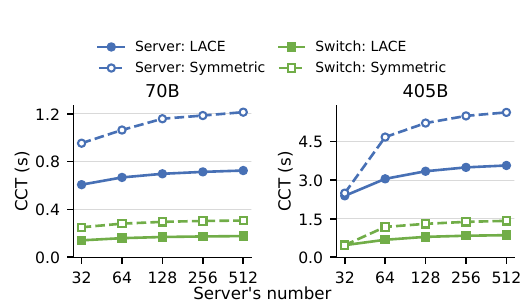}
  \caption{OCS-side CCT under server- and switch-node deployments.}
  \Description{Two panels show absolute CCT in seconds for 70B and
  405B at 32--512 servers. Both deployments benefit from independent
  directional allocation, except for near parity at 32 servers for 405B.
  Switch-node CCT excludes EPS-local communication.}
  \label{fig:app-switch-node}
\end{figure}

\textbf{Switch-node aggregation.}
Figure~\ref{fig:app-switch-node} shows that LACE reduces CCT
relative to symmetric allocation in both deployments.
Server-node deployments show larger absolute time savings,
while switch-node aggregation lowers the OCS-side CCT baseline.
LACE retains substantial relative gains after aggregation,
showing that independent directional allocation remains
beneficial when switches serve as fabric nodes.
Appendix~\ref{app:eps_sensitivity} examines how servers per EPS
and uplink provisioning affect gains when server-access costs are included.

\textbf{Multi-OCS routing and repair.}
We map the main configurations onto a three-stage OCS Clos with
16 servers per leaf, 256 access TX lanes and 256 access RX lanes per
leaf, and 256 middle planes. Bipartite edge coloring constructs paths
for all 368 segments across the 20 LACE and Symmetric LACE configurations.
Both data and port-completion circuits satisfy lane and internal-link
exclusivity. CCT is unchanged under the assumption that the stages
reconfigure in parallel with one fabric-level $\delta$.
Appendix~\ref{app:constrained_routing} gives the edge-coloring
construction and the resource checks applied during repair.

\begin{figure}[t]
  \centering
  \includegraphics[height=8.5cm, keepaspectratio, width=\columnwidth]{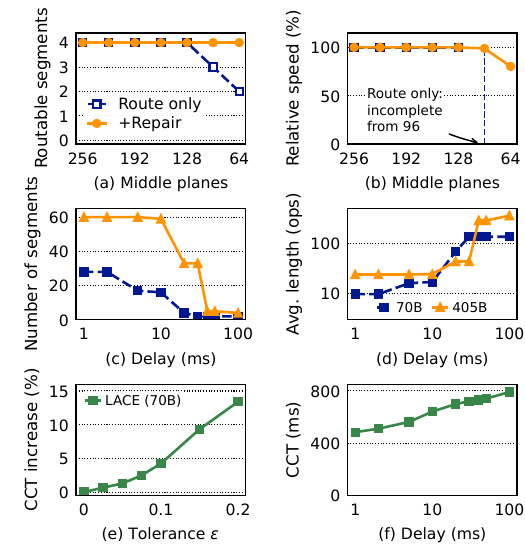}
  \caption{Deployment and sensitivity at 128 servers with $16\times400$~Gbps ports. (a--b) Routing and repair for four 70B segments; speed is
  relative to 256 planes, and routability does not imply meeting
  $\epsilon$. (c--d) Planner segmentation of 70B and 405B Full schedules.
  (e) 70B CCT increase over $\epsilon=0$ at $\delta=20$ ms.
  (f) 70B CCT versus simulated delay at $\epsilon=0.05$.}
  \Description{Six panels show routing success and repaired speed as
  middle planes decrease, segment count and average length as delay
  increases, and the CCT effects of tolerance and reconfiguration delay.
  Repair maps all four segments, retaining 98.85 percent speed at 96
  planes and 80.29 percent at 64. Higher delay produces fewer segments.
  Increasing tolerance from zero to 0.2 raises CCT by 13.40 percent.}
  \label{fig:eval-deployment-sensitivity}
\end{figure}

To test constrained internal connectivity, we reduce middle planes
for the four 70B, 128-server segments
(Figure~\ref{fig:eval-deployment-sensitivity}(a--b)). Route only succeeds for all four
through 128 planes, but only three at 96 and two at 64. Repair preserves
every demanded direction and routes all four at 98.85\% and 80.29\%
of the 256-plane speed. Two segments meet $\epsilon$ after either repair,
compared with three initially. These results validate routing and
repair for the evaluated uniform Clos with leaf-local connections.

\subsection{Segmentation and Parameter Sensitivity}
\label{sec:eval-segmentation}
\label{sec:eval-sensitivity}

\textbf{Higher reconfiguration cost favors longer segments.}
At 128 servers, the 70B and 405B Full schedules contain 274 and 1,445
operations. With $\delta=1$ ms, the Planner forms 28 and 60 segments;
at 20 ms, these fall to four and 33; at 100 ms, to two and four
(Figure~\ref{fig:eval-deployment-sensitivity}(c--d)). These counts precede integer
realization and merging of identical configurations.

\textbf{Tolerance and delay affect different parts of the schedule.}
For 70B Full at 128 servers, raising $\epsilon$ from 0.05 to 0.1
increases CCT by 2.92\%; the increase from zero to 0.2 is 13.40\%
(Figure~\ref{fig:eval-deployment-sensitivity}(e)). Raising $\delta$ from 1 to 100 ms
increases CCT from 0.484 to 0.791 s (63.44\%), while the number of
segments falls from 28 to two. At 30--100 ms, one configuration change
remains, so CCT continues to grow with its cost
(Figure~\ref{fig:eval-deployment-sensitivity}(f)). Appendix~\ref{app:offline_runtime}
reports offline scheduling runtime separately from communication time.

\section{Discussion}
\label{sec:discussion}

\textbf{Dependence on the training schedule.}
LACE exploits directional demand remaining after collective selection
and rank placement; it does not require every operation to be
asymmetric. Sequence parallelism, collective algorithms, and grouping
servers behind an EPS can change both the volume and direction of
OCS traffic. These choices must therefore be reflected in the Demand
Profiler rather than assumed to preserve the reported gains.
LACE currently plans an ordered schedule offline. Changes to that
schedule require replanning, and extending the planner to overlapping
operations or dynamic expert routing requires modeling their readiness
and dependencies. Communication gains may not translate into training speedups under computation overlap.

\textbf{Port inventories and direct connectivity.}
The main deployment targets are multi-port server nodes and switch
nodes, whose ports can provide direct return circuits for feedback.
Single-port nodes with different TX/RX peers would need indirect
return paths, as in prior OCS forwarding designs~\cite{rotornet,opera}.
For software ACKs, intermediate hosts or switches would relay feedback
along a loop-free path, charging every hop's traffic and resources;
without such a path, the node retains a duplex connection.
Port count alone does not determine transport compatibility: both
single- and multi-port servers must deliver feedback to the state
that tracks the transfer, as discussed below.
Likewise, sufficient TX and RX lanes do not guarantee routability
through an arbitrary multi-OCS fabric. Our routing and repair results
cover the evaluated Clos topology, and unresolved mappings require
replanning rather than execution of a partial configuration.

\textbf{Endpoint hardware compatibility.}
Compatibility depends on port capabilities and configuration, rather than NIC generation or whether the endpoint is a NIC, DPU, or EPS. Each must operate with different TX/RX peers under the rate, FEC, and training conditions in \S\ref{sec:endpoint_execution}. A valid RX optical signal alone is insufficient: module loss-of-signal handling and Ethernet local/remote faults can affect transmission~\cite{intel_link_fault,lane_zerwas}. Link-local PAUSE/PFC also requires feedback to reach the correct upstream sender. Our testbed validates its installed NIC, module, and firmware combination, not all such platforms. Drain-and-verify checks transition success; it cannot make unsupported hardware compatible. Such ports require compatible duplex bindings or platform changes.

\textbf{Optical switching architectures.}
3D-MEMS and piezoelectric beam steering permit lane-level allocation
when fibers and controls are independently exposed. Wavelength-selective
switches (WSS) and planar-waveguide networks require checking their
wavelength, path-conflict, and shared-element constraints.
Tunable lasers and arrayed-waveguide gratings (AWGRs) use wavelength
to select destinations~\cite{sirius};
adapting LACE would require wavelength-aware allocation and binding.
Hardware that couples both directions, or an API exposing only duplex
port pairs, cannot directly express our configurations.

\textbf{Reconfiguration delay.}
The Segment Planner charges optical switching, link recovery, and
verification, favoring longer-lived configurations when recovery is slow.
Our prototype observes second-scale recovery (Appendix~\ref{app:testbed}),
as also reported for commodity link initialization by MixNet~\cite{mixnet}.
The millisecond-delay simulations assume correspondingly faster
recovery; the testbed measures recovery separately from fixed-topology
replay and training. Faster recovery would enable shorter communication
phases without changing the scheduling model.

\textbf{Compatibility with RC.}
RC feedback must reach the NIC/QP owning the connection; delivery to
another independent NIC on the same server is insufficient.
RoCEv2 routing permits different forward/return optical paths~\cite{nvidia_rocev2}.
An EPS can forward feedback to the original server NIC without
terminating the QP. A conservative integration would reserve
reciprocal circuits between the same physical port pair, leaving
other lanes independently schedulable. This requires port-binding
constraints; symmetric node-level circuit counts alone are insufficient.
Alternatively, routed return paths could preserve RC endpoints,
subject to routing and transport constraints. Both options require return-path resource accounting
and draining RC traffic before reconfiguration. They remain
unevaluated. Our UC bulk-transfer mechanism does not transparently
implement all RC verbs, including RDMA reads and atomics.

\textbf{Transport scaling and node hardware compatibility.}
Concurrent 400/800~Gbps ports may expose unmeasured CPU limits.
Completion batching, state sharding across cores, and NIC/DPU offload
are candidates; ACK batching alone cannot remove per-chunk processing.
Future NIC interfaces could expose independent data/feedback paths and
a stable transport identity, directing feedback to its reliability
context. Hardware could
handle cumulative ACKs, retransmission, duplicate suppression, and
completion; independent NICs would additionally need explicit state
forwarding or sharing. A quiescence interface could coordinate
reconfiguration with in-flight transfers. Standardizing and implementing
these capabilities could eliminate host-side ACK processing on supported
platforms. LACE would continue to allocate directional capacity and
configuration boundaries; software ACKs provide an execution path
on current hardware.

\section{Related Work}

\textbf{Optical circuit scheduling.}
ACTINA allocates optical resources among predictable TP, DP, and PP
communication domains~\cite{actina}. MixNet adjusts regional
connectivity for dynamic MoE traffic~\cite{mixnet}, while Opus
reconfigures photonic rails at parallelism phase boundaries~\cite{opus}.
RotorNet and Opera instead use traffic-oblivious topology schedules:
RotorNet supports two-hop indirect forwarding~\cite{rotornet}, and
Opera combines multi-hop forwarding for latency-sensitive traffic
with direct circuits for bulk transfers~\cite{opera}.
LACE targets the directed node-pair demand of a selected training
schedule. It jointly chooses which consecutive operations share a
configuration and how many simplex circuits serve each direction,
subject to each node's TX and RX lane inventories.

\textbf{Collective communication and topology optimization.}
NCCL selects collective algorithms and channels~\cite{nccl,demystifyingNCCL};
SCCL and TACCL synthesize topology-aware collective
schedules~\cite{sccl,taccl}; and AdapCC and ResCCL
optimize collective execution and scheduling~\cite{adapcc,resccl}.
RingBiOdd constructs bidirectional AllReduce schedules for odd
MCM meshes~\cite{ringbiodd}; RECCL adapts communication relationships
to optical connectivity~\cite{reccl}; and TopoOpt jointly selects
parallelization, topology, and routing~\cite{topoopt}. These methods
optimize how communication is performed or jointly select its topology.
LACE takes the communication operations and rank placement as inputs, preserving their order while allocating simplex circuits and selecting reconfiguration boundaries.

\textbf{Unidirectional connectivity and link operation.}
Helios explores unidirectional circuits and identifies the
bidirectional assumption in Ethernet fault management~\cite{farrington2010helios}.
Zerwas et al. experimentally examine how optical reconfiguration
interacts with Ethernet link-failure handling~\cite{lane_zerwas}.
RFC~3077 supports unidirectional links through link-layer tunneling
over a separate bidirectional network~\cite{rfc3077}.
These precedents establish that unidirectional connectivity and
separate return paths are not new in themselves. LACE combines
training-aware lane allocation and segmentation with software
acknowledgments over reverse simplex circuits and coordinated
link configuration and recovery.

\textbf{Optical fabric architectures.}
Apollo and Jupiter establish production OCS deployment and topology
engineering~\cite{Apollo,JupiterEvolving}; TPU v4 uses OCS to configure
ML supercomputer partitions~\cite{tpuv4}. LumosCore and InfiniteHBD
provide scalable optical architectures for large AI
clusters~\cite{lumoscore,infinitehbd}. LACE focuses on scheduling
for fabrics exposing independently configurable TX-to-RX connections.
Its Lane Binder maps simplex circuit allocations to physical lanes
and optical paths under the deployed fabric's constraints;
our multi-OCS evaluation considers a three-stage Clos.

\section{Conclusion}
This paper established that pairwise directional skew is a recurring property of training traffic, and that assigning duplex circuits to the two directions of a node pair can leave lanes underutilized. LACE addresses this mismatch through independent TX/RX allocation. It jointly plans configuration boundaries and simplex-circuit allocations offline, then realizes physical connections under node and fabric constraints, without changing the selected collective algorithms, operation order, or rank placement. We separately evaluate software acknowledgments and configuration recovery. With sixteen 400~Gbps ports per server, simulations of LLaMA-3.1 70B and 405B show 1.21--2.04$\times$ communication speedup over ACTINA. Our three-server fixed-topology testbed achieves 1.80$\times$ communication-replay and 1.27$\times$ GPT-2 training speedups over ACTINA.


\bibliographystyle{ACM-Reference-Format}
\bibliography{references}

\appendix
\section{Node-Level Demand and Resource Constraints}
\label{app:planning_formulation}
Let $\mathcal U$ be the set of server nodes or switch nodes defined in
\S\ref{nodes-ports-and-optical-lanes}. Node $u$ contributes $k_u$ TX
lanes and $k_u$ RX lanes. Let $c$ be the per-lane rate in bytes per
second; a rate specified in Gbps is converted before computing time.
For operation $k$, $D^{(k)}=[d_{uv}^{(k)}]$ records directed byte demand,
including the calibrated transport overhead used in the experiments.
Transfers within a node are excluded. For a switch node, this also
excludes transfers between servers attached to the same EPS.
The profiler preserves operation order and does not symmetrize demand.

An optional mask $A_{uv}$ records whether a direct optical path can
exist from $u$ to $v$. For the abstract nonblocking OCS,
$A_{uv}=1$ for all distinct nodes. Positive demand on an unavailable
direction cannot be served directly. The mask expresses reachability,
not simultaneous routability: shared optical-path constraints are
checked by the Lane Binder (Appendix~\ref{app:constrained_routing}).

\section{Fractional Segment Planning}
\label{app:segment_dp}
The allocation $Y_{uv}$ is a fractional \emph{number of simplex circuits},
as in \S\ref{sec:segment_planner}; $cY_{uv}$ is its associated rate.
The fractional feasible set satisfies
\begin{align}
\sum_{v\ne u}Y_{uv}&\leq k_u, &
\sum_{v\ne u}Y_{vu}&\leq k_u, &&\forall u,\label{eq:app_cont_tx}\\
Y_{uv}&\geq0, &
Y_{uv}&=0 &&\text{if }u=v\text{ or }A_{uv}=0.
\label{eq:app_cont_reach}
\end{align}
Each direction demanded in a segment must have positive allocation.
For operation $k$, define
\begin{equation}
L_k(Y)=\max_{d_{uv}^{(k)}>0}\frac{d_{uv}^{(k)}}{cY_{uv}},
\qquad C(I,Y)=\sum_{k\in I}L_k(Y).
\label{eq:app_operation_time}
\end{equation}
A zero-demand operation has zero optical time, and positive demand
with zero allocation has infinite time. This is the sequential-operation
model in Equation~\ref{eq:segment_time}.

\textbf{Continuous reference.}
For fixed segment $I$, minimizing $C(I,Y)$ over the fractional feasible
set is convex: each reciprocal term is convex on positive allocations,
maxima and sums preserve convexity, and the constraints are linear.
An exact solution provides a reference for that segment, not a guarantee
for LACE's heuristic or its subsequent integer realization. Physical
port completion and shared paths impose further constraints.

\textbf{Weighted filling.}
Let $\mathcal D_I$ be the union of demanded directions in $I$.
The implementation initializes weights using
\begin{equation}
w_{uv}^{(0)}=\sqrt{\sum_{k\in I}
\frac{d_{uv}^{(k)}}{\max_{i,j}d_{ij}^{(k)}}},
\qquad (u,v)\in\mathcal D_I,
\label{eq:app_initial_weight}
\end{equation}
with zero-demand operations contributing zero. Small numerical guards
prevent division by zero. Given remaining TX and RX inventories
$r_u^{\mathrm{TX}}$ and $r_v^{\mathrm{RX}}$, each filling round adds
\begin{equation}
\Delta Y_{uv}=\min\left\{
\frac{r_u^{\mathrm{TX}}w_{uv}}{\sum_{j:(u,j)\in\mathcal D_I}w_{uj}},
\frac{r_v^{\mathrm{RX}}w_{uv}}{\sum_{i:(i,v)\in\mathcal D_I}w_{iv}}
\right\}.
\label{eq:app_feasible_fill}
\end{equation}
Source proposals sum to at most the source's remaining TX inventory;
destination proposals do the same for RX. Taking the smaller proposal
therefore preserves both constraints. Inventories are updated after
each round.

The allocator evaluates the original demands and emphasizes transfers
close to each operation's completion time. For nonzero operations, let
$\beta_{uv}^{(k)}=[d_{uv}^{(k)}/(cY_{uv})]/L_k(Y)$.
The next weights are proportional to
\begin{equation}
w_{uv}\leftarrow\sqrt{\sum_{k\in I}
\frac{d_{uv}^{(k)}}{cY_{uv}^{2}}
\bigl(\beta_{uv}^{(k)}\bigr)^8}.
\label{eq:app_bottleneck_reweight}
\end{equation}
Zero entries contribute zero. The exponent gives greater weight to
slow transfers without replacing the max-based evaluation objective.
The allocator performs three rounds of weight updates, each with
twelve filling rounds, and retains the feasible allocation with the
lowest $C(I,Y)$. If each node has at most one outgoing and one incoming
demanded direction, the directions share no TX or RX budgets, and
$Y_{uv}=\min\{k_u,k_v\}$ is optimal for the fractional model.

\textbf{Adjacent merging.}
Starting with one segment per operation, the planner keeps adjacent
merge savings from Equation~\ref{eq:merge_saving} in a priority queue.
After accepting a merge, it updates the neighboring candidates and
discards stale entries. At termination, no evaluated adjacent merge
has positive saving under the heuristic costs; this is not a globally
optimal partition. With fixed candidate-solver effort and caching,
the process evaluates $O(K)$ candidates and uses $O(K\log K)$ queue
work for $K$ operations. Candidate evaluation also depends on the
number of operations and demanded directions in each segment.

\section{Integer Simplex Circuit Realization}
\label{app:discrete_realization}
For segment $I$, let $\mathcal D_I$ contain every direction with positive
demand in at least one operation, including reverse control traffic.
Integer data-carrying circuit counts $X$ satisfy
\begin{align}
\sum_{v\ne u}X_{uv}&\leq k_u, &
\sum_{v\ne u}X_{vu}&\leq k_u, &&\forall u,\label{eq:app_int_tx}\\
X_{uv}&\in\mathbb Z_{\geq0}, &
X_{uv}&=0 &&\text{if }u=v\text{ or }A_{uv}=0,\label{eq:app_int_reach}\\
X_{uv}&\geq1, &&&&(u,v)\in\mathcal D_I.
\label{eq:app_int_coverage}
\end{align}
The performance target is $C(I,X)\leq(1+\epsilon)\widehat C(I)$,
using Equation~\ref{eq:app_operation_time} with integer $X$.
This is a target checked after realization, not a universal
approximation bound.

\textbf{Coverage and additions.}
One circuit per demanded direction is the minimum for direct service.
This assignment fits the abstract OCS exactly when every direction is
reachable and each node's outgoing and incoming peer counts fit its
lane inventories. It need not be physically realizable before port
completion and routing. Further additions follow positive fractional
gaps $Y_{uv}-X_{uv}$ as described in \S\ref{sec:ocs_realizer}.
For a feasible one-circuit addition or removal, the time changes are
\begin{align}
\Delta^+_{uv}(X)&=C(I,X)-C(I,X+E_{uv}),\label{eq:app_add_gain}\\
\Delta^-_{uv}(X)&=C(I,X-E_{uv})-C(I,X),\label{eq:app_remove_loss}
\end{align}
where $E_{uv}$ has a single unit entry. The evaluation implementation
also attempts performance-guided additions using $\Delta^+$ when
target alignment leaves the tolerance unmet and lanes remain.
The reported 70B ablation does not exercise this additional pass.

\textbf{Pruning and outcomes.}
After meeting the target, pruning removes the circuit with the smallest
$\Delta^-$ among removals that preserve coverage and the time tolerance.
Neither additions nor pruning may exceed a TX or RX inventory.
If the search cannot meet the target, it returns its best feasible
allocation with the measured excess time; this does not prove that
no better integer allocation exists. If minimum direct coverage fails,
the segment requires replanning or an explicitly modeled forwarding
alternative. Identical adjacent configurations can share their operation
lists and avoid a reconfiguration; physical bindings must also be
retained for that boundary to incur no switching cost.

\section{Port Completion, Routing, and Connection Reuse}
\label{app:constrained_routing}
\label{app:lane_footprint}
\label{app:physical_binding}
\textbf{Completing enabled ports.}
The commodity-port requirement in \S\ref{sec:lane_binder} applies to
both lanes of each enabled port. Let $Z_{uv}$ count additional simplex
circuits installed solely to complete unused directions of these ports.
The Lane Binder checks the full configuration $H=X+Z$:
\begin{equation}
\sum_v H_{uv}=\sum_v H_{vu}=a_u,\qquad 0\leq a_u\leq k_u,
\label{eq:app_completed_ports}
\end{equation}
where $a_u$ is the number of enabled ports at node $u$. Each of those
ports has exactly one connected TX lane and one connected RX lane;
their peers may differ. The binder must find actual lane assignments
and paths, not merely satisfy these counts. Additional circuits consume
physical resources but contribute no throughput to $C(I,X)$.
Thus the physical circuit footprint is $\sum_{u,v}H_{uv}$, rather
than $\sum_{u,v}X_{uv}$. Peak data-only TX/RX counts provide lower
bounds; they do not by themselves establish a minimum physical port
footprint in a constrained fabric.

\textbf{Path constraints.}
Expand $H$ into individual simplex circuit requests $\mathcal L(H)$.
For request $\ell$, let $\mathcal P_\ell$ contain candidate optical
paths, including their source and destination lanes. Binary choices
$z_{\ell p}$ obey
\begin{align}
\sum_{p\in\mathcal P_\ell}z_{\ell p}&=1,
&&\ell\in\mathcal L(H),\label{eq:app_route_one_path}\\
\sum_{\ell}\sum_{p\in\mathcal P_\ell:e\in p}z_{\ell p}&\leq1,
&&\text{each directional resource }e.
\label{eq:app_route_capacity}
\end{align}
Resources include TX/RX lanes, inter-OCS fibers, and switch input/output
interfaces. Paths must also satisfy each OCS's cross-connect constraints.
Routed connections become usable only after the node-side verification
in \S\ref{sec:endpoint_execution}.

\textbf{Constructive mapping for the evaluated Clos.}
Each leaf serves sixteen server nodes through 256 access TX lanes and
256 access RX lanes. Circuits within one leaf use local connections.
For inter-leaf requests, construct a bipartite multigraph with one copy
of each source leaf on the left and each destination leaf on the right.
Each simplex circuit is an edge. With maximum degree $\Delta$,
bipartite edge coloring uses $\Delta$ colors. Assigning each color to
a middle plane ensures that no two circuits share a leaf's outgoing
or incoming connection to that plane. The mapping succeeds when the
number of planes is at least $\Delta$ and the middle switches support
the resulting matchings.

\textbf{Repair and reuse.}
When routing fails, the binder first tries alternative paths. If circuit
counts must change, repair removes optional data circuits while
preserving every demanded direction, then recomputes port completion
and reroutes the entire $H$. Candidate removals favor smaller increases
in $C(I,X)$; the evaluated Clos repair also checks that excess internal
resource demand does not increase. Recomputing $Z$ is necessary because
removing a data circuit may change which ports need completion.
Each accepted removal decreases $\sum X_{uv}$, so at most
$\sum_{u,v}(X_{uv}-\mathbf1[(u,v)\in\mathcal D_I])$ removals are possible.
Repair may exceed $\epsilon$; it reports the resulting time separately
from routing success. A bounded or heuristic routing failure is an
unresolved mapping, not proof of physical infeasibility.

For consecutive configurations, the binder retains old connections
whose lanes remain in the selected enabled-port sets, then binds
remaining requests. The pairwise quantity
$\min\{H_{uv}^{\mathrm{old}},H_{uv}^{\mathrm{new}}\}$ bounds potential
reuse, but port selection and internal paths can prevent attaining it.

\section{Switch-Node Aggregation and CCT}
\label{app:eps_sensitivity}
\begin{table}[tbp]
\centering
\caption{EPS aggregation. $g$: servers per EPS;
Inter/Excess: percentages of original bytes. Fixed and Full-rate
report access-aware speedup over same-boundary symmetric allocation.
Fixed provides 800~Gbps per EPS per direction; Full-rate provides
$g\times800$~Gbps, matching aggregate server access bandwidth.}
\label{tab:eps_placement}
\footnotesize
\setlength{\tabcolsep}{3pt}
\begin{tabular}{@{}lrlrrrr@{}}
\toprule
Model & $g$ & Mode & Inter & Excess & Fixed & Full-rate \\
\midrule
70B & 4 & DP & 25.0 & 24.6 & 1.750 & 1.000 \\
 &  & PP & 14.3 & 8.8 & 1.000 & 1.000 \\
 &  & Mixed & 41.7 & 32.9 & 1.554 & 1.038 \\
 &  & Full & 42.4 & 33.2 & 1.545 & 1.037 \\
\addlinespace[2pt]
 & 8 & DP & 12.5 & 12.3 & 1.750 & 1.000 \\
 &  & PP & 0.0 & 0.0 & 1.000 & 1.000 \\
 &  & Mixed & 32.0 & 13.7 & 1.173 & 1.038 \\
 &  & Full & 32.8 & 14.3 & 1.179 & 1.039 \\
\midrule
405B & 4 & DP & 25.0 & 24.6 & 1.750 & 1.000 \\
 &  & PP & 20.0 & 11.1 & 1.283 & 1.000 \\
 &  & Mixed & 47.4 & 37.0 & 1.493 & 1.040 \\
 &  & Full & 47.8 & 37.3 & 1.498 & 1.042 \\
\addlinespace[2pt]
 & 8 & DP & 12.5 & 12.3 & 1.750 & 1.000 \\
 &  & PP & 6.7 & 3.7 & 1.000 & 1.000 \\
 &  & Mixed & 31.5 & 26.6 & 1.593 & 1.002 \\
 &  & Full & 31.6 & 26.6 & 1.591 & 1.002 \\
\bottomrule
\end{tabular}
\end{table}

Table~\ref{tab:eps_placement} reports a separate analysis of earlier
70B/405B traces on 128/512 servers, with 800~Gbps server access,
100~Gbps circuits, $\delta=20$ ms, and $\epsilon=0.05$.
Placement minimizes cross-EPS bytes plus unmatched directional bytes.
Each operation costs the larger of its access and optical times,
plus reconfiguration between segments; local transfers retain access
cost. These are model estimates, not EPS hardware measurements.

Aggregation reduces crossing traffic, but its benefit depends on
placement and uplink provisioning. Full-rate uplinks can shift the
bottleneck to server access, leaving little gain despite residual
optical asymmetry; more servers per EPS do not imply monotonically
smaller gains. Direct-connect fabrics such as TopoOpt~\cite{topoopt}
avoid EPS aggregation and retain server-level directional demand.

\begin{figure}[tbp]
    \centering
    \includegraphics[width=\columnwidth]{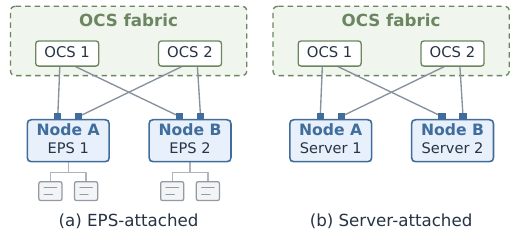}
    \caption{Node and fabric boundaries in two deployments. Tabs mark
    OCS-facing ports; optical links represent TX/RX fiber pairs.}
    \Description{Two parallel OCSes form the optical fabric in each
    panel. Nodes are EPSes with attached servers in panel (a), and
    directly attached servers in panel (b).}
    \label{fig:node_fabric_mapping}
\end{figure}

\section{Offline Runtime and Scalability}
\label{app:offline_runtime}
We measure runtime on one CPU core of a Linux x86-64 server using
Python 3.10, with BLAS/OpenMP restricted to one thread. Each of the
20 model/scale/port configurations has one warm-up and three timed
repetitions. The measured stages are the Segment Planner, Circuit
Realizer, port completion with a deterministic reference binding,
and a subsequent binding pass that retains existing connections.
The total includes both binding passes. Input loading, serialization,
demand profiling, and multi-OCS routing are excluded. Totals are summed
within each repetition before taking the median.

\begin{figure}[t]
  \centering
  \includegraphics[width=\columnwidth]{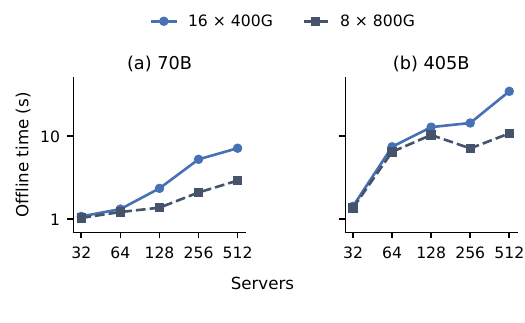}
  \caption{Offline scheduling time on one CPU core. Lines show
  medians and bands min--max over three repetitions; totals include
  planning, realization, port completion, and the binding passes
  described above.}
  \Description{Two panels compare sixteen 400G ports with eight 800G
  ports for 70B and 405B. At 512 servers, sixteen-port totals are
  7.10 and 34.06 seconds, respectively.}
  \label{fig:eval-runtime}
\end{figure}

At 512 servers, measured total time is 7.10 s for 70B and 34.06 s
for 405B with 16 ports at 400G. With eight ports at 800G, the times
are 2.91 and 10.67 s (Figure~\ref{fig:eval-runtime}). More integer
circuits increase the search work, but runtime also depends on segment
structure and need not grow monotonically with cluster size. These
times are offline planning costs, separate from the simulated CCT.

\section{Physical Testbed Measurement Details}
\label{app:testbed}
\textbf{Communication replay and training.}
Both experiments use OCS topologies on three servers, each with
two 100G ports. Replay uses two complete runs per topology, a
native three-server mapping, and TCP transfers with real acknowledgments.
Per-port rate limits are 10~Gbps for replay and 5~Gbps for training,
with the same striping policy across topologies. These limits keep
traffic below the host processing ceiling and isolate the effect of
which peers the lanes connect to.

Training uses one RTX 5000 and two RTX 4000 GPUs, sequence length 1024,
microbatch size one, accumulation eight (global batch 24), FP16, and
AdamW. Each topology runs one epoch from the same pretrained checkpoint
and data order. Wall time includes all optimizer steps and excludes
validation and checkpoint saves. Validation uses 96 fixed packed blocks;
the maximum paired training-loss difference is 0.002984.

\textbf{Software acknowledgments and CPU cost.}
The microbenchmark uses a fixed 100G link with data and feedback on
the same NIC port and host-memory buffers. Each message size has
30 paired RC/UC trials, with randomized method order and 32 messages
per trial. Figure~\ref{fig:eval-software-ack} reports the mean of
$100(t_{\mathrm{UC+ACK}}/t_{\mathrm{RC}}-1)$ with a paired 95\% bootstrap
interval, and median throughput. Both modes use 64-KiB chunks and
a 128-chunk window. UC batches cumulative ACKs every eight chunks,
with transfer-boundary flushing and a 10-$\mu$s timer.

Table~\ref{tab:ack_cpu} reports the median, over 30 trials per mode
and size, of the sum of sender and receiver process CPU time divided
by application GiB transferred. Both endpoints use one busy-polling
thread, with median CPU-time/wall-time ratios above 0.996 in all
groups. There is no additional ACK thread. These costs include
polling and all transport processing; similar values do not imply
zero ACK cost or establish spare CPU capacity. The experiment does
not measure concurrent multiport processing or GPU-memory transfers.

\begin{table}[tbp]
  \centering
  \caption{Combined endpoint CPU cost on the fixed 100G link.
  Values are median CPU-s/GiB, including busy polling.}
  \label{tab:ack_cpu}
  \begin{tabular}{lrr}
    \toprule
    Message size & RC & UC + software ACK \\
    \midrule
    64 KiB & 0.4697 & 0.5674 \\
    1 MiB  & 0.2033 & 0.2067 \\
    16 MiB & 0.1769 & 0.1772 \\
    64 MiB & 0.1759 & 0.1760 \\
    \bottomrule
  \end{tabular}
\end{table}

Large transfers still require chunk-level processing: a 64-MiB
message contains 1,024 chunks. Ignoring wire overhead, retransmissions,
and timer/boundary ACKs, a payload rate $R$ in bits/s requires
$R/(8\cdot65{,}536)$ chunks/s and one progress-triggered ACK per
eight chunks. At 400/800~Gbps this is approximately 0.763/1.526
million chunks/s per port; at 6.4 Tb/s it is 12.207 million chunks/s
per node in one direction, for either port configuration.
These are processing-demand estimates, not measured CPU costs or
predictions of the required core count.

\textbf{Configuration verification and recovery.}
A separate control experiment performs three triangle-to-ring and
three ring-to-triangle transitions with 100G configured, autonegotiation
disabled, and the existing FEC configuration preserved. Each transition
checks port status and reception of configuration-ID probes on all
intended connections. All six pass; controller-observed readiness is
5.79--7.44 s. This interval includes control commands, polling, and
probe verification. In one missing-RX fault injection, verification
fails, one port-reset attempt does not restore the missing path, and
the controller restores and verifies the preceding topology. The
observed 40.65-s interval includes two 15-s timeouts and recovery actions.
Original topology and port settings are restored after the experiment.

The recovery test is unloaded and excludes application draining and
RDMA QP recreation. As also observed by MixNet~\cite{mixnet}, current
commodity link initialization can dominate optical switching time.
The simulation delay is a separate fabric parameter; deployment uses
the full delay exposed to communication, including link recovery and
verification.

\section{Per-Lane Training Payload Estimates}
\label{app:ack_payload_estimate}
The shaded range in Figure~\ref{fig:eval-software-ack} is an analytical
comparison with the measured message sizes. We use a Megatron configuration~\cite{megatron-lm}
with TP=8 within each server, sequence parallelism enabled, context
parallelism disabled, sequence length 8,192, and microbatch size one.
These assumptions are specific to this estimate; the main scheduling
experiments retain their original generated communication schedules.
Traffic for a direction is assumed to be evenly striped across
$\ell=16$ TX lanes, with rank/channel traffic sharing a sustained stream.
Using fewer lanes increases the estimated bytes per lane proportionally.

\textbf{DP Ring rounds.}
For DP group size $D$, use the default target bucket size
$B=\max\{40{,}000{,}000,1{,}000{,}000D\}$ parameter elements per rank.
One logical Ring round for a full bucket sends $B/D$ elements per rank.
Aggregating eight ranks and striping across $\ell$ lanes gives
\begin{equation}
M_{\mathrm{DP}}=\frac{8Bb}{D\ell}\quad\text{bytes per lane per round},
\label{eq:app_dp_payload}
\end{equation}
where $b=2$ for BF16 AllGather and $b=4$ for FP32 ReduceScatter.
A complete Ring collective has $D-1$ rounds; the estimate does not
combine those dependent rounds into one message. Tail buckets may
be smaller, and actual bucket sizes depend on parameter grouping.

\textbf{PP transfers.}
With sequence parallelism, each TP rank transfers one eighth of the
activation tensor at a PP boundary. For sequence length $S$,
microbatch size $m$, and hidden dimension $h$, the eight ranks together
send $Smh$ BF16 elements. Hence
\begin{equation}
M_{\mathrm{PP}}=\frac{2Smh}{\ell}\quad\text{bytes per lane per transfer}.
\label{eq:app_pp_payload}
\end{equation}
For $h=8192$ (70B) and $h=16384$ (405B), this gives 8 and 16 MiB
per lane. A backward activation-gradient transfer has the same volume
under the same shape and BF16 precision. The 1F1B order changes when
these transfers occur, not their individual size.

\begin{table}[htbp]
\centering
\caption{Full-bucket DP payload per lane per Ring round, in MiB.
Entries show BF16 AllGather / FP32 ReduceScatter with sixteen TX lanes.
PP transfers are 8 MiB/lane for 70B and 16 MiB/lane for 405B.}
\label{tab:app_lane_payload}
\small
\begin{tabular}{@{}rcc@{}}
\toprule
Servers $N$ & 70B ($D=N/8$) & 405B ($D=N/16$)\\
\midrule
32  & 9.54 / 19.07 & 19.07 / 38.15\\
64  & 4.77 / 9.54  & 9.54 / 19.07\\
128 & 2.38 / 4.77  & 4.77 / 9.54\\
256 & 1.19 / 2.38  & 2.38 / 4.77\\
512 & 0.95 / 1.91  & 1.19 / 2.38\\
\bottomrule
\end{tabular}
\end{table}

Table~\ref{tab:app_lane_payload} gives the range across the evaluated
model scales. The 0.95--38.15 MiB band covers these full-bucket DP
rounds and the two PP sizes; it is not a percentile interval. These
are aggregate payloads per lane, which can consist of multiple NCCL
channels and transport chunks.

\clearpage
\end{document}